\documentclass[fleqn,usenatbib]{mnras}
\usepackage[T1]{fontenc}
\usepackage{ae,aecompl}

\usepackage{graphicx}	
\usepackage{amsmath}	
\usepackage{amssymb}	

\title[First IR QPO in a black hole X-ray binary]{Detection of the first infra-red quasi periodic oscillation in a black hole X-ray binary}

\author[Kalamkar et al.]{
M. Kalamkar,$^{1}$\thanks{E-mail: maithili86@gmail.com} P. Casella,$^{1}$ P. Uttley,$^{2}$ K. O'Brien,$^{3}$ D. Russell,$^{4}$ T. Maccarone,$^{5}$ 
\newauthor M. van der Klis$^{2}$ and F. Vincentelli$^{1,6,7}$
\\
$^{1}$INAF, Osservatorio Astronomico di Roma, Via Frascati 33, I-00078 Monteporzio Catone, Italy\\
$^{2}$Astronomical Institute Anton Pannekoek, University of Amsterdam, Science Park 904, 1098XH Amsterdam, Netherlands\\
$^{3}$Department of Physics, University of Oxford, Keble Road, Oxford OX1 3RH, UK\\
$^{4}$New York University Abu Dhabi, PO Box 129188, Abu Dhabi, UAE\\
$^{5}$Department of Physics, Texas Tech University, Box 41051, Lubbock, TX 79409, USA\\
$^{6}$Università degli Studi dell'Insubria, Via Valleggio 11, I-22100 Como, Italy\\
$^{7}$INAF - Osservatorio Astronomico di Brera Merate, via E. Bianchi 46, I-23807, Merate, Italy
}

\date{Accepted XXX. Received YYY; in original form ZZZ}

\pubyear{2015}

\begin{document}
\label{firstpage}
\pagerange{\pageref{firstpage}--\pageref{lastpage}}
\maketitle

\begin{abstract}
We present analysis of fast variability of Very Large Telescope/ISAAC (infra-red), \textit{XMM-Newton}/OM (optical) and EPIC-pn (X-ray), and RXTE/PCA (X-ray) observations of the black hole X-ray binary GX 339-4 in a rising hard state of its outburst in 2010. We report the first detection of a Quasi Periodic Oscillation (QPO) in the infra-red band (IR) of a black hole X-ray binary. The QPO is detected at 0.08 Hz in the IR as well as two optical bands (U and V). Interestingly, these QPOs are at half the X-ray QPO frequency at 0.16 Hz, which is classified as the type-C QPO; a weak sub-harmonic close to the IR and optical QPO frequency is also detected in X-rays. The band-limited sub-second time scale variability is strongly correlated in IR/X-ray bands, with X-rays leading the IR by over 120 ms. This short time delay, shape of the cross correlation function and spectral energy distribution strongly indicate that this band-limited variable IR emission is the synchrotron emission from the jet. A jet origin for the IR QPO is strongly favoured, but cannot be definitively established with the current data. The spectral energy distribution indicates a thermal disc origin for the bulk of the optical emission, but the origin of the optical QPO is unclear. We discuss our findings in the context of the existing models proposed to explain the origin of variability.
\end{abstract}



\section{Introduction}\label{intro}
\noindent The emission from the accretion flow and jets in the hard state of black hole X-ray binaries (BHB) spans from the radio band to $\gamma$-rays. The accretion flow around the black hole has two main components - a geometrically thin optically thick accretion disc (modelled as a quasi-blackbody at low X-ray energies) and a geometrically thick optically thin inner hot flow/magnetized corona (Comptonized emission, modelled by a power-law component at high X-ray energies). The blackbody disc emission component is relatively weak in the hard states (as opposed to the soft states) and the accretion flow spectrum is dominated by the Comptonized emisson \citep[e.g., see][]{done2007}. The jets consist of energetic particles accelerated away from the black hole, and its synchrotron emission extends from radio through optical infra-red (OIR) and possibly into X-rays \citep[e.g.,][]{markoff2001,fender2009}. The emission in ultraviolet (UV) and OIR may also include intrinsic thermal emission from the outer region of the accretion disc and reprocessing of X-rays incident on the outer accretion disc \citep[e.g.,][]{paradijs-1994}. The fluxes in various bands shows correlations e.g., radio/X-rays \citep{corbel-2003-rad-x,gallo-2003-rad-x}, X-rays/optical infra-red (OIR) \citep{homan-339-multi-2005,russell-2006-oir-xr,coriat-339-multi-2009}. Clearly, the accretion flow and jets form a strongly connected and interacting complex system which is also observed to undergo dramatic changes due to mass accretion rate variations during episodic outbursts. \\
\\
To comprehend fully the relationship between the accretion flow and the jet during an outburst, it is crucial to disentangle the contribution to emission from various components. The synchrotron radio emission is interpreted to be from a self-absorbed compact jet. The compact jet models predict that above a certain frequency, commonly referred to as the jet spectral break frequency (expected in the OIR range), the jet is not self-absorbed and produces optically thin emission \citep{blandford-jet-1979}. This makes the OIR bands of great interest as both the accretion flow and jet can be detected in these bands. Despite the limitations due to high extinction and contribution from the accretion disc and/or the companion star, spectral modelling of multi-band datasets which include OIR data in bright hard states has led to detection of the jet spectral break in the mid-IR bands in a handful of sources \citep[see e.g.,][]{corbel-2002-339-break, gandhi-339-midirbrk-2011,russell-2013-spec-brk}. \\
\\
X-ray emission from the accretion flow in the hard state is known to vary on slow (tens of seconds) and fast (sub-second) time scales \citep[e.g.,][]{klis2006} and often narrow peaked components called Quasi Periodic Oscillations (QPOs, classified type-C, \citeauthor{wijnands1999qpos} 1999; \citeauthor{casella2005} 2005) are observed in the power spectrum. Until now only a few examples of QPOs have also been seen in the UV/optical band, at either similar or commensurate frequencies as those in the X-ray band, and their origin has been unclear. Reprocessing of band-limited X-ray variability within the accretion flow and/or on the companion star has been observed in some sources allowing us to put constraints on the size of the accretion disc/system \citep[see e.g.,][]{brien-echoes-2002}. Many results later showed that not all variability can be accounted for by simple reprocessing scenarios. There is growing evidence of fast variability arising in the synchrotron jet emission in the OIR bands, and that this jet variability is correlated with X-ray variability \citep{motch-339-1983, kanbach-1118-2001, hynes-1118-2003, gandhi-339-2010, casella-irvar-jet-2010,lasso-2013-1915-xrircorr}. This variability correlation provides a valuable tool to probe the accretion flow-jet interaction.\\ 
\\
\noindent GX 339-4 is a recurrent black-hole X-ray binary transient located at a distance of $>$ 6 kpc \citep{hynes-339-dist} with a mass function of 5.8 solar masses and an orbital period of 1.75 days \citep{hynes339}. The source has exhibited several outbursts in the past and it entered a new outburst on January 3, 2010 \citep[MJD 55199;][]{yamaoka-339-2010-atel}. In this paper, we report the study of simultaneous multi-wavelength observations taken on March 28 2010 (MJD 55283) in infra-red, optical/UV and X-rays when the source was in the hard state during the rise of the outburst. We study variability in each of these bands along with the broad-band spectral energy distribution (SED). The aim is to investigate the origin of variable emission and its correlation amongst different energy bands to probe the accretion flow-jet coupling. We describe the observations and data analysis in Section \ref{section:obs-data-ana}. The variability and SED results and their comparison with previous reports are presented in Section \ref{section:results}. The origin of variable emission is discussed in the context of different models in Section \ref{section:discussion} with summary and conclusions in Section \ref{section:summ-conc}.
\section{Observations and Data Reduction}\label{section:obs-data-ana}
\subsection{Infrared data}
\noindent GX 339-4 was observed between MJD 55283.270 and MJD 55283.361 using Very Large Telescope/Infrared Spectrometer And Array Camera \citep[ISAAC;][]{moorwood-1998-isaac}. The observation was taken with the Ks (effective wavelength at 2.2 $\mu$m) filter with a useful exposure of about 4 ksec. ISAAC was operated in FastJitter mode, allowing us to store each detector integration time (DIT) as a slice in a FITS cube. Each slice has a DIT of 37 ms and there are 995 slices per cube. We obtained 165 cubes with a short gap between each cube for file merging and fits header writing. Each slice is a 256 $\times$256 pixel ($\sim38''\times38''$) image containing the target, a bright reference star and a faint comparison object. We have extracted light curves of the three sources using the ULTRACAM data reduction pipeline. Each source was extracted using standard aperture photometry based on parameters derived from the reference star position and profile. 
\subsection{Optical/UV data}
\noindent The \textit{XMM-Newton} Optical Monitor \citep{mason-xmm-om-2001} observed GX 339-4 (Obs. ID 0654130401) in the fast mode allowing a high time resolution of 500 ms. The data was obtained in eight exposures typically 3 ksec each employing the V filter (effective wavelength at 543 nm) for the first three exposures and U filter (effective wavelength at 344 nm) for the consequent five exposures. The second and third exposures with U filter were simultaneous with the IR data. Here we analyse these two exposures and the last exposure with the V filter, which is closest to the IR dataset. The data was analysed using the SAS pipeline \texttt{omfchain} script with the default set of parameters to obtain a light curve for each exposure.
\subsection{X-ray data}
\subsubsection{\textit{XMM-Newton}}
\noindent We analyse EPIC-pn \citep{struder-epic-2001} timing mode data (time resolution 0.03 ms) of the \textit{XMM-Newton} observation (Obs. ID 0654130401) of GX 339-4. The observation is about 33.5 ksec long, but we use only the data strictly simultaneous with the IR data (about 4 ksec). The data is reduced with the standard procedure with XMM SAS (v14.0.0). The raw events were processed using SAS tool \texttt{epchain}. The source events were extracted with RAWX columns 28 to 48 as the source is centred at RAWX column 38. We test for pile-up using \texttt{epatplot} and find the data to be piled-up. Hence, we remove the central pixel and obtain the light curves in multiple energy bands (see below). The background was low and without any flaring events.
\subsubsection{Rossi X-ray Timing Explorer}
\noindent The Rossi X-ray Timing Explorer (\textit{RXTE}) observed the 2010 outburst of GX 339-4. We use the observation (Obs. ID 95409-01-12-01) which is simultaneous with the IR observation (about 3 ksec). We reduce the Proportional Counter Array \citep[PCA,][]{jahoda2006} data following standard procedures described in the \textit{RXTE} Cookbook using HEASOFT v6.16. The PCA has five Proportional Counter Units out of which at most two were switched on during the observation. We use only the  data simultaneous with the IR data and when two PCUs were switched on. We exclude data in which the elevation (angle between Earth's limb and the source) is less than 10$^{\circ}$ or the pointing offset (angle between source position and pointing of the satellite) exceeds 0.02. We obtain light curves with the Event mode data (which has a time resolution of 122 $\mu$s) in multiple energy bands (see below). 
\section{Results}\label{section:results}
\subsection{Power spectra}\label{section:pds}
\noindent We obtain the power spectral densities  (PSD) by averaging the modulus square of the discrete Fourier transform of uninterrupted light curve segments. The PSD is the variance of signal at each Fourier frequency, providing us a measure of variability of the signal on various time scales. The Poisson noise is estimated from the errors on the count rate and subtracted \citep[see e.g.,][]{gandhi-339-2010}.  The PSDs are normalised such that their integral power gives the squared rms variability. We then fit the PSD with multiple Lorentzian profiles.\\
\\
The time resolution of the IR data is 37 ms, giving a Nyquist frequency of 13.51 Hz. The light curve had gaps (between 6 seconds to 10 seconds) shorter than the uninterrupted segments (typically 37 seconds or 26 seconds). In order to explore the low frequency variability with high frequency resolution, these gaps were filled with count rate values normally distributed around a mean (and the  standard deviation) calculated from the preceding and following light curve segments of half the gap length. To study the effects of this on the PSD, we perform two tests. With the observed IR PSD, we simulate light curves using the method of \cite{timmerkoenig1995}. Gaps identical to the IR light curve were artificially introduced in the simulated light curves and the resulting PSD is compared to the PSD of the original light curve. In the second test,  we introduce gaps in the RXTE X-ray light curve (which does not have any gaps) identical to the simultaneous IR light curve, fill these gaps using similar technique as in the IR data, and compare the PSDs. In both cases, no artificial features were observed in the simulated light curve PSD and its shape is similar to the original PSD. The integrated fractional rms amplitudes of the PSDs generated with light curves with and without their gaps filled differ by only 2\% (due to the additional noise from the filled gaps). We also investigated other methods to fill the gaps (e.g., linear interpolation) and found similar results.  \\
\begin{table}
\center
{\renewcommand{\arraystretch}{0.65}
\begin{tabular}{|c|c|c|}
\hline 
Energy band & Frequency (Hz) & QPO frac. rms (\%)\\
\hline
IR & 0.080 $\pm$ 0.001 & 6.0 $\pm$ 1.0\\
Optical V & 0.081 $\pm$ 0.002 & 8.3 $\pm$ 1.4 \\
Optical U & 0.082 $\pm$ 0.002 & 4.4 $\pm$ 1.0 \\
X-ray 0.3-2 keV & 0.092 $\pm$ 0.004 & 5.3 $\pm$ 1.7\\
X-ray 2-60 keV & 0.096 $\pm$ 0.004 & 5.8 $\pm$ 1.4\\
\hline
X-ray 0.3-2 keV & 0.160 $\pm$ 0.009 & 10.1 $\pm$ 2.7\\
X-ray 2-60 keV & 0.161 $\pm$ 0.003 & 11.6 $\pm$ 1.5\\
\hline
\end{tabular}}
\caption{Best fit QPO parameters in various energy bands shown in Figure \ref{fig:pds}. Errors are at 1 standard deviation. } \label{table:para}
\end{table}

\begin{figure}
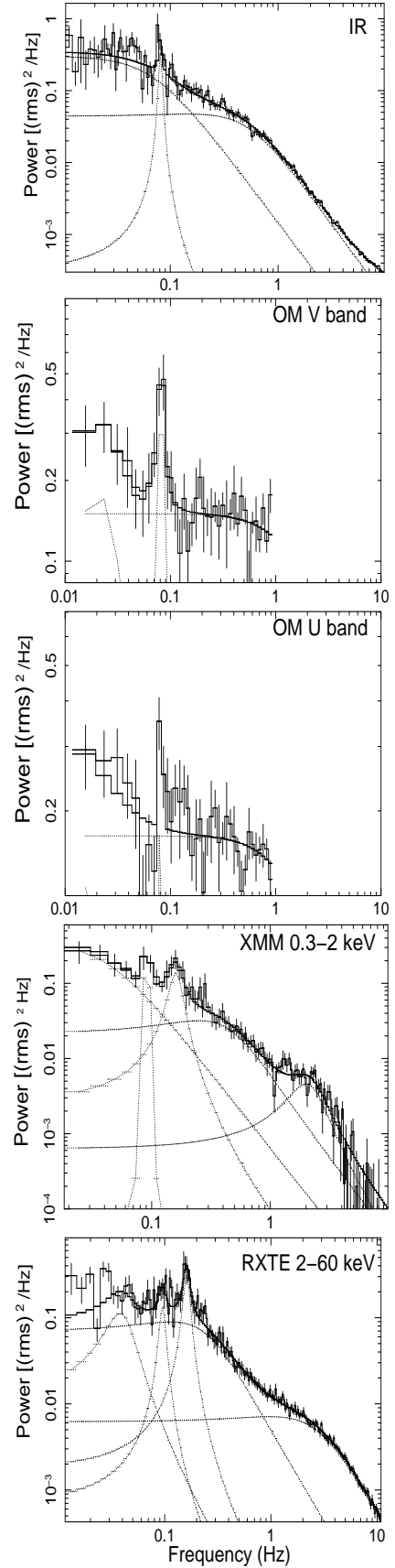

\center
\includegraphics[width=4.5cm,height=5.5cm,angle=-90]{irpsd-lc2full-16384-1.05-lorns.ps}
\includegraphics[width=4.5cm,height=5.5cm,angle=-90]{404-v-pds.ps}
\includegraphics[width=4.5cm,height=5.5cm,angle=-90]{405-406-u-pds.ps}
\includegraphics[width=4.5cm,height=5.7cm,angle=-90]{xmm-0.3-2pds-wg.ps}
\includegraphics[width=4.5cm,height=5.5cm,angle=-90]{rxte2-60pds.ps}
\caption{Power spectra in the various energy bands as indicated. The best fit model using multiple Lorentzians to each Poisson noise subtracted power spectrum is shown. Except for the OM V band, all power spectra have simultaneous data.}\label{fig:pds}
\end{figure}

\noindent The IR PSD is fit with three Lorentzian profiles which gives a reduced $\chi^2$ value of 1.48 (for 162 \textit{dof}), which is shown in Figure \ref{fig:pds}. We report, for the first time in the IR band, a Quasi Periodic Oscillation (QPO) in the PSD of a BHB. It is significantly detected (4.5 $\sigma$) at 0.080 $\pm$ 0.001 Hz and is accompanied by two band-limited noise components. The significance is calculated as the ratio of the power to negative error; the negative error is calculated by varying the model parameter such that $\Delta\chi^2$=1. Similar band-limited noise components were reported when the source was in persistent low hard state more than a year after its outburst in 2007 with strong indications of the IR emission originating from the jet \citep{casella-irvar-jet-2010}.\\
\\
The OM PSDs in the U and V bands are obtained using the same method as above, except the data did not have gaps. The time resolution of the data is 500 ms, giving a Nyquist frequency of 1 Hz. The best fit was obtained with three Lorentzian profiles with a reduced $\chi^2$ value of 1.005 (for 31 \textit{dof}) and 0.775 (for 31 \textit{dof}) for the U and V bands, respectively as shown in Figure \ref{fig:pds}. The QPO is detected significantly in the U and V bands respectively at 0.082$\pm$0.002 Hz and 0.0812$\pm$0.002 Hz (at 3.25 $\sigma$ and 4.12 $\sigma$, respectively). We note that instrumental effects on PSD have not been investigated for the OM, so there might be an uncertainty on the Poisson level. In addition, due to a high background in this observation, the fractional rms amplitudes we report in Table \ref{table:para} should be taken with caution. However, our main result is that the optical QPO frequency is consistent with the QPO frequency in the IR band, which is not altered by the instrumental effects. A QPO in the optical band has been reported before in this source in previous outbursts \citep{motch-opqpo-1982, gandhi-339-2010} and in other sources such as XTE J1118+480 \citep{hynes-1118-2003} and Swift~J1753.5-012 \citep{durant-1753-2009}. The shape of the PSD at low frequencies is also similar to the one reported in the 2007 post-outburst state \citep{ gandhi-339-2010}. Their data allowed a higher Nyquist frequency (10 Hz) compared to our data (1 Hz) and band-limited noise was reported in that data which peaked around 1 Hz.\\

\noindent The PSD in the 0.3-2 keV (soft) X-ray band and the 2-60 keV (hard) X-ray band (and multiple sub-bands reported in Figure \ref{fig:sed}) were obtained respectively with the XMM-Newton and the RXTE data similar to the method described for the optical data. The PSD were fitted with five Lorentzian profiles giving a reduced $\chi^2$ value of 1.37 (for 109 \textit{dof}) and 1.20 (for 163 \textit{dof}) in the soft and hard bands, respectively. The PSD are shown in Figure \ref{fig:pds}. Band-limited noise variability is accompanied by a QPO, which is typical of hard state PSDs. The QPO is detected at 0.164$\pm$0.003 Hz (3.1$\sigma$) and  0.161$\pm$0.003 Hz (5.8$\sigma$), in the soft and hard bands. The X-ray QPO has been classified as the type-C QPO and is commonly observed in many BHB. It has been detected and traced in all of GX 339-4 outbursts \citep[e.g.,][]{belloni2005, motta339}. We also detect another QPO close to the sub-harmonic of the type-C QPO; it is detected in the soft and the hard bands respectively at 0.092 $\pm$ 0.004 Hz (2.2 $\sigma$) and 0.096$\pm$0.004 Hz (2.97 $\sigma$) which is close to the IR QPO frequency. The X-ray sub-harmonic QPO has been previously reported in this source \citep[e.g.,][]{belloni2005}. The frequency of the X-ray sub-harmonic QPOs in this source and other sources has been observed with a scatter around the expected frequency i.e., at half the type-C QPO frequency \citep[see e.g.,][]{pawar-2015-qpo-harm-mod}, similar to what we observe here; the origin of this scatter remains unknown.   \\

\noindent The detection of an IR QPO is a significant result as we now have detection of QPOs from X-rays to IR bands, encompassing a broad multi-wavelength range. However, what is more interesting is the harmonic relationship between the frequencies at which the QPOs are detected in various bands. We note that in our data, the IR and optical QPO are at half the X-ray QPO frequency, closer to the sub-harmonic of the X-ray type-C QPO. Similar behaviour has also been reported before in optical and X-ray bands during a rising hard state along an outburst \citep{motch-339-1983}. It is interesting to note that QPOs  detected in another BHB XTE J1118+480 simultaneously in optical, UV and X-ray bands were at the same frequency \citep{hynes-1118-2003}. Hence, with only a handful of multi-band detections in few sources, we already see a varied behaviour of the QPO.
\begin{figure}
\center
\includegraphics[width=8.cm,height=8.cm,angle=-90]{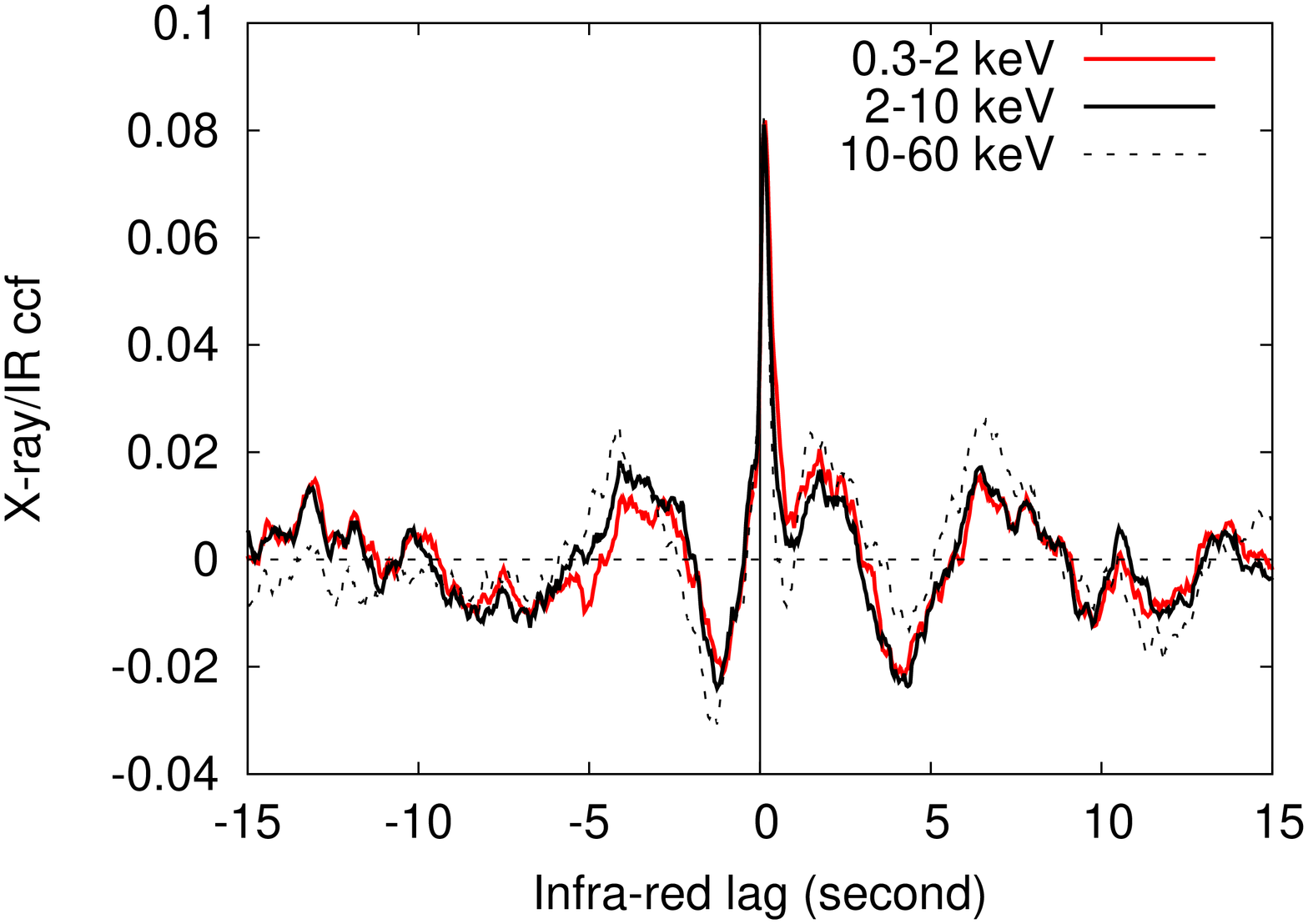}
\includegraphics[width=5.cm,height=8cm,angle=-90]{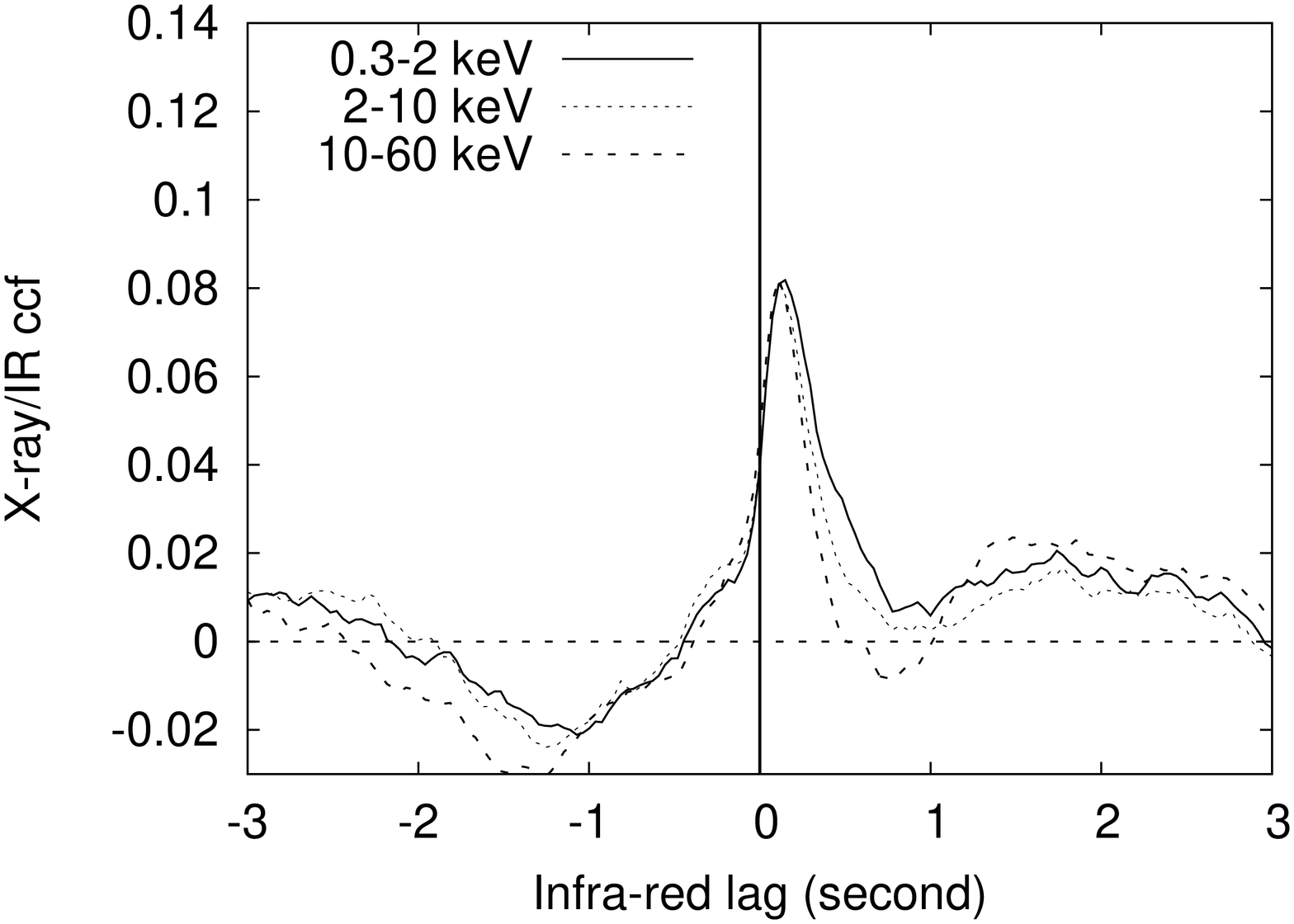}
\caption{Cross correlation function showing the lag in IR band against the different X-ray bands (as indicated) is shown in the top panel and the zoom-in view in the bottom panel. The vertical line shows zero lag.}\label{fig:ccf}
\end{figure}
\subsection{Cross correlation functions}\label{section:ccfs}
\noindent The barycentered IR and X-ray light curves are uniformly binned at the time resolution of IR band of 37 ms and the gaps are excluded. The hard X-ray band is subdivided in two bands - 2-10 keV and 10-60 keV. We inspect the cross correlation function (CCF) between the IR and various X-ray bands, which are shown in Figure \ref{fig:ccf}. In the top panel, the X-ray versus IR CCF is shown, and its zoom in view is shown in the bottom panel; a positive lag indicates delay of the IR band with respect to the X-ray band. The lags reported here are the weighted average of the CCF above the half maximum. The errors on the lags were estimated using the bootstrap technique. An average CCF and its weighted lag were calculated from randomly selected light curve segments from the original light curve (segments were allowed to repeat). This process was repeated 10$^3$ times and the dispersion of these weighted lags gives the error on the lag. We associate an additional systematic offset (due to the IR CCD read-out) of 2 ms on all lags. As the time resolution of our optical data is 500 ms, we cannot investigate optical lags on short time scales and hence do not report the optical lags here. \\
\\
The IR band lags the 0.3-2 keV band by 153$\pm$3 ms, the 2-10 keV band by 128$\pm$4 ms and the 10-60 keV band by 121$\pm$3 ms; IR lag against the soft 0.3-2 keV band is reported for the first time. In the case of the origin of IR emission due to reprocessing on outer disk or the companion star, longer lags and a highly asymmetric CCF  \citep[due to different light travel times to the disc;][]{brien-echoes-2002} are expected. The shape of the CCF we observe is nearly symmetric (particularly in the hard bands). This argues against reprocessing and points towards a jet origin (also see Section \ref{section:origin-bbn}). The broad modulation humps at longer lags are on the time scales of the X-ray QPO period ($\sim$ 6 s). The difference in the IR lag against the soft 0.3-2 keV and hard X-ray bands (which are consistent with each other within errors) can be explained as the soft/IR band CCF peaks at a slightly longer lag (148 ms compared to 111 ms lag against both the hard bands; which are limited by the time resolution of 37 ms) and has a slight asymmetry towards longer lags; this leads to a longer weighted average lag. The shape and lag of the soft/IR band CCF is a result of contribution from the disc (see below) to this band, compared to the hard band emission from a relatively compact hot flow which is located closer to the jet. \\
\\
The only other X-ray/IR CCF reported before was also for this source in a lower luminosity hard state where it showed an IR delay of $\sim$ 100ms with a broad and nearly symmetric CCF \citep{casella-irvar-jet-2010}; no broad modulations were seen due to the absence of the QPO. They also rule out reprocessing and favour synchrotron jet emission based on similar arguments.

\begin{figure}
\center
\includegraphics[width=6.5cm,height=9cm,angle=-90]{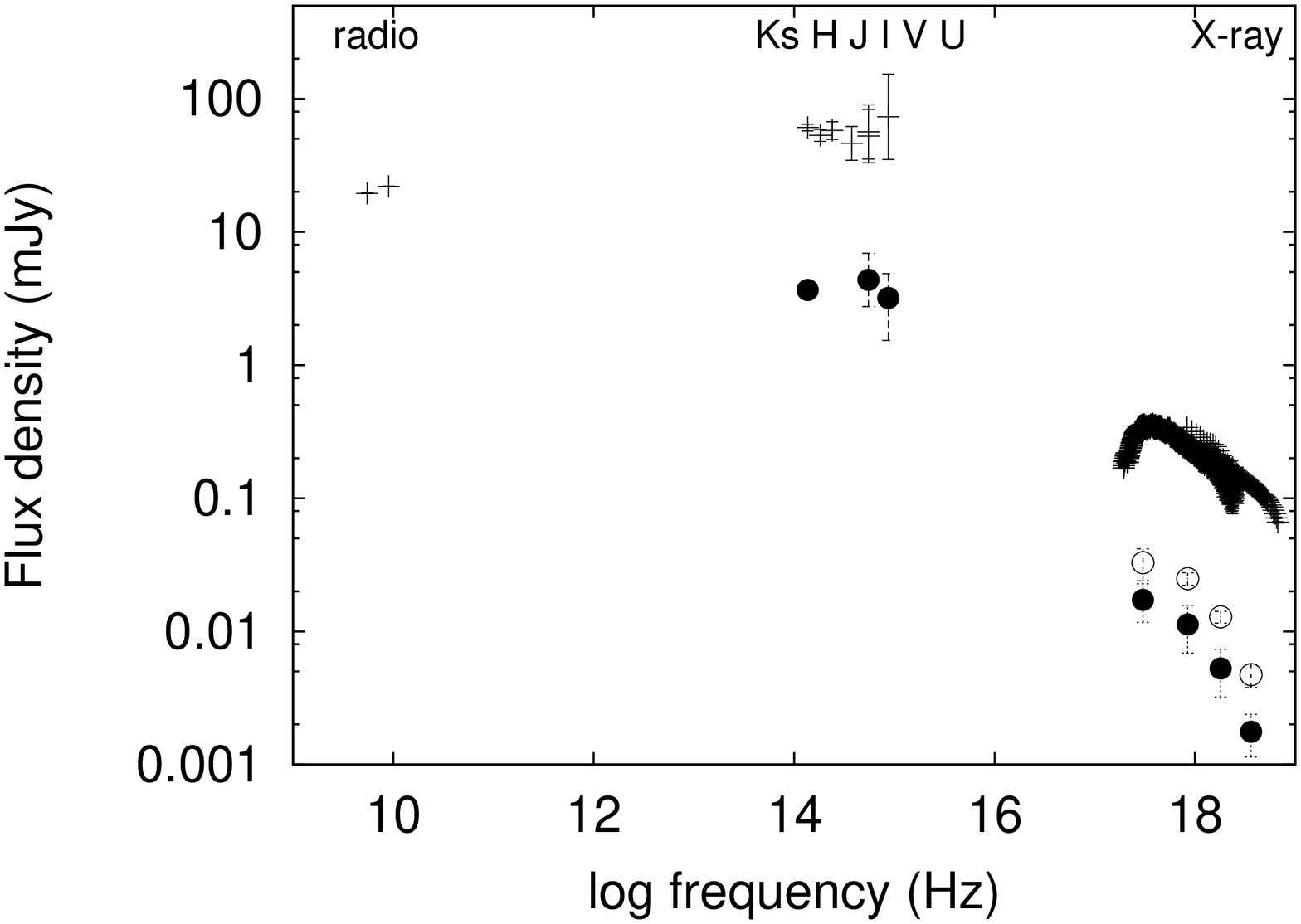}
\includegraphics[width=6cm,height=9cm,angle=-90]{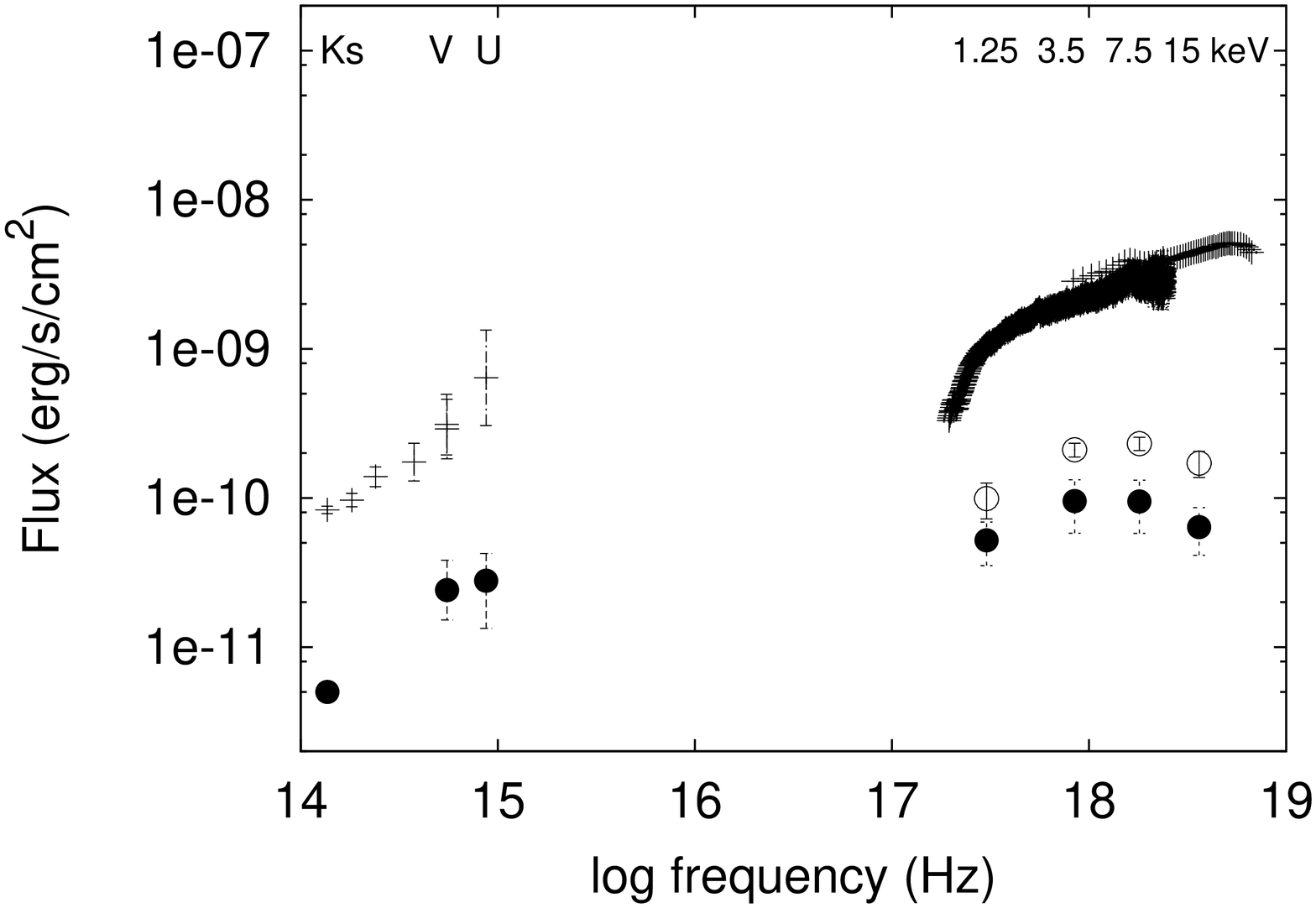}
\caption{\textit{Top panel}: Broad band spectral energy distribution spanning radio with ATCA \citep{corbel-2013-339}, Ks band with ISAAC, H,J,I,V bands with SMARTS \citep{buxton-339-2012} and V,U bands with XMM-Newton OM and X-rays with XMM-Newton and RXTE. The XMM-Newton, RXTE, ISAAC and OM data are simultaneous, while the other band data are closest to these observations. The spectrum of the QPO is shown with filled circles (QPOs around 0.08 Hz) and empty circles (QPOs at 0.16 Hz) in the bands as indicated in the bottom panel. \textit{Bottom panel}: The QPO spectrum in flux units in the bands 0.5--2 keV, 2--5 keV, 5--10 keV, and 10--30 keV where the mid-point of X-ray bands is shown, alongwith the mean spectrum. The errors on the OIR QPO and mean spectrum are dominated by deredenning uncertainty.}\label{fig:sed}
\end{figure}
\subsection{Broad band spectral energy distribution}\label{section:sed}
\noindent Figure \ref{fig:sed} top panel shows the broad band SED from radio to X-rays. The most nearly simultaneous radio observations were on the same day obtained with Australia Telescope Compact Array (ATCA) reported by \cite{corbel-2013-339-rad-x}. The IR data are from our ISAAC observation. The source was observed with the 1.3m SMARTS telescope \citep{subasavage-2010-smarts} in V, I, J and H bands \citep{buxton-339-2012}. There are multiple measurements taken within 0.1 day of the IR observation in the V, I and J bands and the light curves show very little variability on that day; hence we use the mean magnitude on that day to obtain the fluxes. The closest H band observation is available from the day before. The U and V band fluxes from the OM data are also represented here. The de-reddened IR and optical fluxes are obtained using standard interstellar extinction law \citep{cardelli-1989-extinction} with A$_{\rm V}$ of 3.25 \citep{gandhi-339-midirbrk-2011}. The X-ray fluxes were obtained using the  XMM-Newton EPIC-pn and  RXTE PCA data. The spectra are obtained using the standard procedure as described in  \cite{plant-339-2015} and \cite{cadolle2011-339}, respectively. The data are jointly fit with the model \texttt{phabs*(diskbb+gaussian+powerlaw)} and an overall constant to account for cross calibration. We obtain a reduced $\chi^2$ of 1.2 (1933 dof) which is sufficient for our purpose. We do not fit the broad band SED with a physical model as this is beyond the scope of this paper and will be presented elsewhere.  \\
\\
The radio spectral index  $\alpha$ (taking a definition of  $S_{\nu}\propto\nu^{\alpha}$, where $S_{\nu}$ is the spectral density at the frequency $\nu$) was reported to be 0.24 which is commonly seen for the optically thick synchrotron emission of a compact jet \citep{corbel-2013-339-rad-x}. The I, V and U band data indicate a positive slope (limited by the large uncertainty due to dereddening) suggesting a thermal disc component. Thermal disc emission has been suggested to contribute in the UV/optical bands in observations obtained few days before our observations \citep{gandhi-339-midirbrk-2011,buxton-339-2012,rahoui-339-multi2012}. Since these observations, the X-ray and optical fluxes increased till our observation \citep{buxton-339-2012}, and hence we expect reprocessing to dominate in the optical bands. The companion star is weak and its contribution can be neglected here \citep{motch-339-1983,shahbaz-2001-339-vlt}. In addition to a thermal disc, there have also been suggestions of its origin in pre-shock jet synchrotron (\citeauthor{gandhi-339-2010} 2010; however see \citeauthor{buxton-339-2012} 2012 for a discussion of dereddening effects) and from a hot accretion flow \citep[][see below]{veledina-op-xr-rep2011}. \\
\\
The Ks, H and J band fluxes fall below the slope from radio data and do not lie on the extrapolation of the thermal component with the expected positive slope, and is hence most likely synchrotron emission from the jet. Variability in the Ks band on time scales of at least 128 ms (at 3 \% fractional rms amplitude) observed in the PSD puts an upper limit of $\sim$ 4 $\times$ 10$^9$cm on the size of the emitting region. With a flux of 3 \% of the average IR flux (8.3 $\times$ 10$^{-11}$ ergs$^{-1}$cm$^{-2}$), we derive a minimum brightness temperature of about 2.5 $\times$ 10$^7$ K. The thermal X-ray flux in the 2-10 keV band from such an optically thick region would be of the order of $\sim$ 0.58 ergs$^{-1}$cm$^{-2}$. Clearly this enormous flux level is ruled out by the data. Lack of data below the Ks band frequency makes it difficult to constrain the jet break, and hence the Ks band could be emission from below or above the break frequency.\\
\\
We also present the first rms spectrum of the QPO ranging from X-rays to IR band which is shown in Figure \ref{fig:sed}. The QPO and the mean spectrum are shown in flux density units in the top panel, and flux units in the bottom panel. The filled circles are the QPOs around 0.08 Hz and the empty circles are the X-ray QPOs at 0.16 Hz. As discussed above, the mean spectrum shows that the IR band originates in the jet and the optical bands from the disc; however, the QPOs in these bands are observed at the same frequency pointing towards a same origin. The errors on the OIR QPO fluxes are large due to large dereddening uncertainties, but the slope of the QPO spectrum appears to be flatter relative to the mean spectrum (which also has the same uncertainties due to dereddening), again suggesting a same origin. The X-ray QPO shows a hard spectrum similar to the mean X-ray spectrum; similar behaviour of the QPO (but with a slightly harder QPO spectrum) in this source was also reported by \cite{axelsson-2015-339-qpo-harm-spec}. However, the rms spectrum of the sub-harmonic has not been investigated in this source, which we report here to be similar to the QPO spectrum.   
\section{Discussion}\label{section:discussion}
The outburst of GX 339-4 in 2010 was a target of wide multi-wavelength coverage and the overall behaviour was similar to its previous outbursts and very typical of BHBs \citep{cadolle2011-339, rahoui-339-multi2012, nandi2012-339,corbel-2013-339}. \cite{gandhi-339-midirbrk-2011} reported a jet spectral break in the MIR range on MJD 55266. Our SED is observed a few days later on MJD 55283 where the optical and IR fluxes are higher. Although a positive correlation between the jet break and luminosity is expected, there is no concrete evidence for this \citep[see e.g.,][]{russell-2013-spec-brk}. Hence, at a higher luminosity the break might be at/below the Ks band, but we cannot constrain the exact location of the break from our data.  \\
\\
We observe band-limited noise variability and a QPO in the OIR and X-ray bands where a correlation is observed between IR and X-ray bands. Due to the lack of high time resolution data in the optical bands we cannot investigate the origin of fast variability in these bands, and its correlation with IR and X-ray bands. However, from the SED the optical band emission seems to be dominated by the thermal disc component. The origin of optical emission in terms of a reprocessed component will be presented in future works. From the X-ray variability studies, there are strong indications that the band-limited noise and the more coherent QPO have different driving mechanisms \citep[see e.g.,][]{ingram-lt-2011}. Hence, we investigate the origin of OIR variability in a similar context.
\subsection{Origin of band-limited noise variability}\label{section:origin-bbn}
The PSD in X-ray and IR bands show strong variability over three decades of frequency. The X-ray emission is from the geometrically thick optically thin hot flow, with some contribution from the accretion disc in the soft band. The origin of the X-ray band-limited noise has been attributed to fluctuations in the mass accretion rate \citep{lyub1997}. These fluctuations arise at different radii in the accretion flow with faster ones generated at smaller radii; slow fluctuations are generated at larger radii and can propagate to smaller radii \citep[see e.g.,][]{uttley-2001}. The shape of the IR PSD is very similar to the X-ray PSDs. The question to address here is what drives the IR variability, if it is connected to X-rays, and if not, what other processes are at play to generate the IR variability independent of the X-rays. \\
\\
The SED suggests that the IR band emission is most likely the synchrotron jet emission. However, there are other processes that could also provide additional contribution to the emission in the IR band and perhaps also the \textit{variability} in this `bright' hard state. The binary separation in this system is $\sim$ 25 light seconds \citep{gandhi-339-2008}. The short IR lag of around 150 ms in the CCF rules out X-ray reprocessing on the companion star. Such a short time scale for reprocessing of X-rays on the outer regions of the accretion disc would imply a highly inclined disc. A high inclination angle has been shown to be unlikely due to absence of absorption dips and eclipses in the X-ray light curves \citep[e.g.,][]{plant-339-2015}. In the outer regions of the geometrically thin accretion disc from where the IR emission can originate, variability is expected to be on viscous time scales which would be of the order of days, as opposed to the much shorter time scales we observe in the PSD.\\
\\
\cite{veledina-op-xr-rep2011} proposed that a geometrically thick optically thin hot flow which gives rise to X-ray emission could also be a source of optical/IR emission. They consider a hybrid flow, where inner regions of the flow give Comptonized X-ray emission from the thermal population of electrons, while the non-thermal electrons in the outer regions give synchrotron OIR emission. The spectral results suggest that the disc is truncated not far from the black-hole \citep{plant-339-2015}; however a hot flow extending upto 100 gravitational radii is required to emit in the IR band. Moreover, in this scenario, we also expect the IR variability to lead the X-rays, as opposed to the observed IR lag. The fast X-ray variability arising in the inner regions of the flow cannot drive the IR variability in the outer regions, as the fluctuations cannot propagate outwards to larger radii \citep{chu2001}, as also noted by them. \\
\\
The only other region where the IR emission is generated in the BHB is from the jet. Hence, we conclude that the band-limited \textit{variable} IR emission is most likely synchrotron emission from the jet. This was also reported before in this source by \cite{casella-irvar-jet-2010} in a low intensity hard state. We now report similar IR variability in a brighter hard state, where the average IR flux is at least four times higher. \citeauthor{malzac-jet-var-2014} 2014 (and references therein) proposed a driving mechanism for jet IR variability. In their model, jet emission is produced by electrons accelerated in internal shocks which are driven by fluctuations of the jet velocity. The fluctuations in the Lorentz factor follow the X-ray PSD of the accretion flow. This model naturally predicts jet IR variability and its correlation with the X-ray variability. Using the X-ray power spectrum reported by \citeauthor{gandhi-339-midirbrk-2011} 2011 (where they observe the jet break), \cite{drappeau-jet-var2015} demonstrated that the spectral and variability properties of the jet IR emission can be explained. Hence, this is a potential model that can provide a driving mechanism for the jet IR variability we observe.
\subsection{Origin of the QPO}
\subsubsection{X-ray QPO}
Many models have been proposed for the origin of QPOs in X-ray bands \citep{stella1998, titarchuk1999, wagoner2001, ingram2009, shaposhnikov-qpo-2012, varniere-qpo-harm2012,veledina-qpo-lt-2013,veledina-op-repo-qpo-2015} which invoke either general relativistic effects or instabilities in the accretion flow. A geometrically thick optically thin tilted hot flow around a spinning black hole undergoes Lense-Thirring precession, giving rise to the X-ray QPO \citep{stella1998,fragile2007, ingram-lt-2011}.  This model can also explain many properties of the QPO such as the frequency evolution along the outburst and its rms spectrum \citep{ingram-lt-2011, ingram2012-frame-dragging}. Recent works show strong indications that the QPO has such a geometrical origin \citep{ingram2012-frame-dragging, motta-2015-qpo-geom, heil-2015-geom}, unlike the band-limited noise variability. 

\subsubsection{Optical QPO}
Models that can consistently explain the QPOs in different energy bands remain sparse. The origin of the \textit{variable} optical emission is unclear, hence we discuss all possibilities for completeness. The IR and optical QPO are observed at the same frequency, which strongly suggests a common emission mechanism. \cite{veledina-qpo-lt-2013} proposed that a hot flow, where inner regions give Comptonized X-ray emission while the outer region emit synchrotron optical emission (as discussed in the previous section), can undergo Lense-Thirring precession and give rise to optical/IR and X-ray QPOs. Although the IR emission from the hot flow is ruled out (as it requires an extended hot flow, see above), a smaller hot flow (about 30 gravitational radii) can still emit in the optical band and provide a mechanism to generate optical QPO. The fundamental frequency is the QPO frequency observed in the optical band. This mechanism was suggested to drive the optical QPO observed in the BHB SWIFT J1753.5-012, where there is a hint of an X-ray QPO at the same frequency \citep{veledina-2015-opqpo-1753}. The model further predicts that the QPO fractional amplitude in optical is greater than X-rays, while X-rays may have much stronger harmonics than the optical. From Table \ref{table:para}, it can be seen that the V band and IR band QPO have a higher fractional rms amplitude than in X-rays at the V band QPO frequency (however keeping in mind the uncertainty over optical QPO \textit{rms} as discussed in Section \ref{section:pds}). We don't detect any harmonics in OIR, with an upper limit of 2.7 \% in the IR band, 3.7 \% in the V band and 2.5 \% in the U band for a QPO at 0.16 Hz, which is lower than the amplitude of the X-ray QPO at 0.16 Hz. Hence, the frequencies and amplitudes of the optical QPOs compared to the X-ray QPOs we observe are in line with the predictions for this model.\\
\\
The model also predicts that the X-ray QPOs can exhibit harmonics stronger than the fundamental for a range of inclination angles (45$^{\circ}$--75$^{\circ}$, and depending on the observer's azimuth). GX 339-4 has been suggested to have an inclination angle in that range \citep[see e.g.,][and references therein]{motta-2015-qpo-geom, heil-2015-geom}, suggesting that the strongest X-ray QPO could actually be the harmonic frequency and hence the optical QPO is at the fundamental frequency. However, the amplitudes of the X-ray QPOs we observe, where the fundamental (close to 0.08 Hz) is weaker than the harmonic (at 0.16 Hz), are much higher than their predicted values which lie below 3 \% \citep[see Figure 5 in][]{veledina-qpo-lt-2013}. Also as discussed in the previous section, the causal connection between the band-limited IR and X-ray noise cannot be explained by this model. \\
\\
Another mechanism was proposed by \cite{veledina-op-repo-qpo-2015}, where the hot flow emits X-rays and undergoes Lense-Thirring precession illuminating varying outer regions of the accretion disc and generates a reprocessed optical QPO. It is limited by the ratio of the QPO period to the light travel time to the illuminated outer disc region. In GX 339-4, their model expects an approximate critical optical QPO frequency of 0.055 Hz for a 10 solar mass BH. A calculation with a lower mass BH renders a higher critical frequency of 0.313 Hz for the optical QPO. This is a potential mechanism for the optical QPO and will be investigated in future works. However, the IR QPO would be generated at larger radii and will be smeared out and hence, not observable.   

\subsubsection{IR QPO}
We investigate the origin of the IR QPO as a geometrical effect, similar to the X-ray QPO. Origin of the   IR QPO from the hot flow via its Lense-Thirring precession or due to reprocessing on the outer regions of the accretion disc can be ruled out as discussed in the previous section. Doppler beaming effects from a precessing jet can provide the mechanism to generate QPOs. Such a mechanism has been proposed in blazars, where periodic flaring is associated with beaming effects in a precessing jet \citep[see e.g.,][]{ abraham-2000-qpo-blazar,caproni-2013-bllac-jetpre, king-2013-qpo-blazar}, and also in tidal disruption events \citep[see e.g.,][]{wang-2014-qpo-tde}. A precessing jet driven by the black hole spin would indicate precession of the black hole spin axis, which is extremely unlikely \citep{nixon-2013-jet-prec}. If the accretion disc drives the jet, a precessing jet would indicate the precession of the accretion disc (or part of it close to the black hole). The time scale on which the full disc would precess is much longer \citep[order of weeks;][]{maloney-1997-diskprec} than the IR QPO time scale we observe. However, the Lense-Thirring precession of the hot flow which is used to explain X-ray QPOs would allow the jet to precess on the QPO time-scale if it is anchored to the hot flow, and we consider this the most likely origin of the IR QPO.  \\ 
\\
If the jet precesses at the same frequency as the hot flow, the X-ray and IR QPOs are expected to be observed at the same frequency, and/or (depending on the specific model), harmonics of the same frequency. The strongest X-ray and IR QPOs are harmonically related in our case which suggests the possibility that the strongest X-ray QPO (the 0.16 Hz QPO here) is the harmonic and not the fundamental QPO. As discussed in the previous section, an X-ray harmonic stronger than its fundamental has been suggested due to inclination effects \citep{veledina-qpo-lt-2013}. In such a case, the hot flow as well as the jet would have the same precession frequency and can explain the apparent harmonic relation between the IR and X-ray QPO. \\
\\
We estimate the possible amplitude of the QPO in an extremely simplistic scenario. The Doppler boosted flux for a compact jet with a spectral index of $\alpha$ is given as $S_{obs}=S_o D^{2-\alpha}$, where D is the Doppler boosting factor given as $D=1/\gamma(1-\beta cos \theta)$ ($\gamma$ is the Lorentz factor, $\beta$=v/c and $\theta$ is the angle of the jet with observer's line of sight). Assuming a mildly relativistic jet with $\beta$=0.7 and a precession angle of 1$^{\circ}$ around an inclination angle of 45$^{\circ}$, the ratio of fluxes when the jet is pointing more or less towards us is $S_{obs+}/S_{obs-}$=$1.071$ \citep[e.g.,][]{fender2003-jetmotion} for a flat spectrum with $\alpha$=0, giving an amplitude of about 6 \%. Strongly relativistic jet will give stronger modulation, and a combination of increasing the inclination and the precession angles for a range of spectral indices gives similar or higher amplitudes. Hence we conclude that, given the most likely origin of aperiodic variable IR emission in the jet, it is plausible that the IR QPO also originates from the jet, possibly via precession of the hot flow which anchors the jet and produces the X-ray QPO.\\

\section{Summary and conclusions}\label{section:summ-conc}
This work highlights the importance of simultaneous broad band multi-wavelength spectral and variability  study. We report the spectral and variability properties of the BHB GX 339-4 from its outburst in 2010 in the X-ray, optical and IR bands in the hard state. We summarise our results below:\\
1. We report the first detection of a QPO in the IR band for a BHB. The QPO is detected in the hard state at a frequency of 0.080$\pm$0.001 with a fractional rms amplitude of 6.0$\pm$1.0 \%.\\
2. A QPO at same frequency is also detected in the optical U and V bands. These OIR QPOs are at a sub-harmonic frequency of the type-C QPO detected in the X-ray bands.\\
3. The band-limited IR variability lags X-ray variability by more than 120 ms. Such a short lag and the shape of the SED rule out reprocessing and suggest that the IR emission and the variability is most likely jet emission. \\
4. As the IR band is dominated by jet emission, the IR QPO can originate in the jet, where we consider jet precession tied to the Lense-Thirring precession of the hot flow. However, more evidence is required to confirm this.\\
\\
The QPO detections reported earlier in X-ray to optical bands can now be extended to the IR band. The harmonic relation between these QPOs provide strong constraints for the QPO models. More multi-wavelength observations are required to observe if this harmonic relation is unique to GX 339-4, or also observed in all/some sources. It is also important to know if the QPO harmonic relation between multiple bands is observed throughout the outburst, or only during a part of it. Hence, to establish a QPO model, an ensemble of such detections is essential. These results demonstrate that variability study is an extremely powerful tool to probe the accretion flow - jet connection, providing strong constraints on the origin of emission in various bands and hence on the structure and geometry of BHBs.\\
\\
\section*{Acknowledgements}
We thank the VLT, XMM and RXTE schedule planners for their successful efforts in scheduling these simultaneous observations. PC and MK thank European Union FP7 Career Integration Grant ``MultiFast" (CIG 322259) for support. MK thanks Chris Done and Luigi Stella for useful discussions. MK thanks Riccardo Campana for software support.









\bsp	
\label{lastpage}
\end{document}